# The Hydrophobic Interaction Induced Strengthening of Hydrogen Bond in Water-DMSO Binary Mixture

**Sangita Mondal and Biman Bagchi\***

*SSCU, Indian Institute of Science, Bangalore 560012, India.*

*\*Email: bbagchi@iisc.ac.in; profbiman@gmail.com*

## Abstract

The lifetime of a hydrogen bond between water and dimethyl sulfoxide (DMSO) is found to be *considerably longer* than that between two water molecules in the neat water. This is counter-intuitive because the charge on the oxygen in DMSO is considerably less than that in water. Additionally, the strength of the water-dimethyl sulfoxide (w-D) hydrogen bond is found to be strongly composition dependent; the lifetime of the hydrogen bond is ten times larger at 30% over that at very low concentration. Using computer simulations, we perform microscopic structural and dynamic analysis to find that these anomalies arise at least partly from an "action-at-a-distance" effect where the attraction between the hydrophobic methyl groups results in self-aggregation of DMSO molecules that "cages" both rotational and linear motions of molecules involved. This is reflected in the observed strong correlation of the lifetime with the local coordination number of the associated methyl groups. The elongated w-D h-bond lifetime causes a slowdown of collective dynamics and affects the lifetime of the w-w h-bond. This nonlinear feedback mechanism explains the strong composition dependence of viscosity and anticipated to play a dominant role in many self-assemblies. Furthermore, the w-D hydrogen bond breaking mechanism changes from low to high DMSO concentration, a phenomenon not anticipated *a priori*. We introduce a new order parameter-based free energy surface of the bond breaking pathway. A two-dimensional transition state rate theory (TSRT) calculation is performed for the lifetime of the w-D h-bond that is found to be semi-quantitatively accurate.

# I. Introduction

Hydrogen bond strength and stability are sensitive to the environment and are determined by many-body interactions, which, for water, are still not fully understood. The typical estimates of bond energy of HB bonds in water range from 3-5 Kcal/mole.[1] Two hydrogen-bonded water molecules in liquid water are held together by an extended network, thus enhancing the enthalpic contributions.[2–4] The network serves to stabilize the bond while the latter in turn is responsible for the network.[5–7] This synergy plays an important role in many natural and biological systems.[8–13] It is however hard to estimate in theory this effect because structure and dynamics are intimately coupled. A measure of the dynamic stability can be obtained by studying the lifetime of a tagged hydrogen bond.

The strength and nature of hydrogen bonds have been subjects of discussion in other non-aqueous systems. Desiraju and Stein have introduced the concept of weak hydrogen bond that appears to be omnipresent in natural systems.[7] In many cases, a system such as a self-assembly is stabilized by a large number of such bonds. What is often not realized or stated explicitly is that the structure so produced also stabilizes the bond itself. The cooperative interplay among various interactions, such as hydrophobic and hydrogen bonding,[14] lies at the heart of biological complexity. Thus, there is a synergistic effect which serves to lower the entropy and enhance the enthalpic stabilization.

Water-DMSO binary mixture finds extensive applications across various fields such as cell biology, cryoprotection, pharmacology, and more.[15–19] The intrigue surrounding this system stems not only from its distinctive chemical and biological applications and functions, but also from the diverse characteristics exhibited by the water-DMSO binary mixture, both as a solvent and a reaction medium. In the seminal work by Luzar and Chandler [20] it was suggested that at 1:2 DMSO-water composition (1DMSO:2$H_2$O) ratio leads to a stable structure. Rousseau and coworkers[21] reported that there is a reduction in the local tetrahedral order of water molecule as the DMSO concentration increases up to $X_{DMSO} = 0.35$. They proposed that the anomalous behavior observed at $X_{DMSO} = 0.35$ could be attributed to the distortion of the local environment of water molecules, leading to an increase in local rigidity. Numerous physical properties of DMSO/water mixtures, including viscosity,[22,23] density,[24] heat of mixing,[25] and diffusion rate,[26] show maxima or minima around 30–40 mol% DMSO.[27] Theoretical methods and calculations have been widely employed to anticipate the structure and configurations of hydrogen bonds (HBs). At low concentrations, DMSO is anticipated to create stable 1DMSO:2H$_2$O species. In mixtures where DMSO predominates (>50%), an association between DMSO molecules with 2 DMSO : 1 water complex is supported, whereas in water-rich mixtures, a tetrahedral ordering of water molecules is induced.[28,29]

Crowding and heterogeneity are fundamental characteristics of molecular biology, particularly at the cellular level.[30–32] The interior of a cell is densely populated, with biomacromolecules or osmolytes confining as much as 20-40% of intracellular water. This crowding phenomenon disrupts solvent hydrogen bond networks and influences the hydration levels of proteins and osmolytes.[33,34] However, quantifying these effects poses challenges, as it necessitates a combination of structural sensitivity and chemical specificity. Consequently, microscopic understandings of complex mixtures needed to correlate solvent environments with biomolecular structure are still in their early stages of development. A crucial initial step in this endeavour involves characterizing the impact of crowding on hydrogen bonding. This lays the foundation for achieving a comprehensive microscopic description of heterogeneous environments.

Because of the many-body interactions, one can obtain insight from studies of the lifetime of a hydrogen bond as a measure of stability and strength. An interesting criterion of hydrogen bond lifetime was introduced by Stillinger[35] and further developed by Luzar and Chandler by using Heaviside functions.[36] Two types of lifetime functions were proposed. We call them $C_I(t)$ and $S_C(t)$, defined below. Most calculations find the decay of both the functions is exponential, with time constant equal to 0.54 ps and 4.5 ps, respectively. In order to understand the reaction free energy surface, we imagine a two-dimensional free energy where the O---O distance and the libration angle serve as the two-order parameters. We can consider the variation of free energy along these two coordinates. In water, there are thus two clear bond-breaking or reactive motions: one is the oxygen-oxygen distance, and the other is the libration angle. The two-dimensional free energy surface will thus have many minima. with multiple exit routes binary mixtures between water and amphiphilic solutes have often produced anomalous results which are hard to understand because of the complexity of interactions. This elegant picture of the hydrogen bond-breaking mechanism gets altered in the case of hydrogen bonds in complex systems, like aqueous binary mixtures.

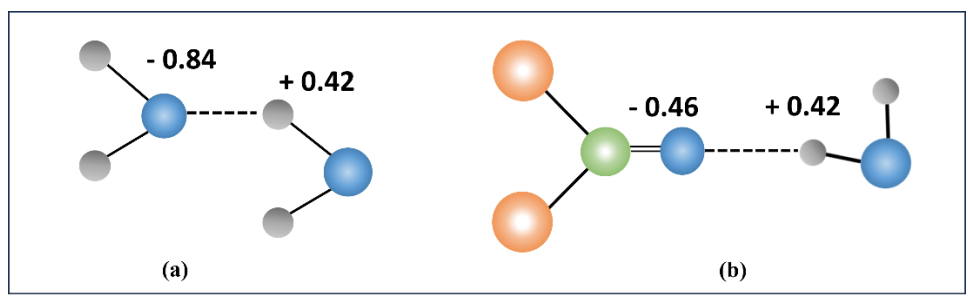

**Figure 1. A schematic representation of (a) water-water (w-w) hydrogen bond, (b) water-DMSO (w-D) hydrogen bond. The blue, grey, green, and orange spheres represent the oxygen atom, hydrogen atom, sulfur atom, and methyl group, respectively. The intermolecular hydrogen bonds are shown by black dashed lines.**

Our analysis reveals a picture somewhat different from the conventional one. We find that there are two aspects that leads to a different picture. First, the composition dependent viscosity peaks sharply near $X_{DMSO}=0.38$, not at $X_{DMSO}=1/3$, as the one DMSO hydrogen bonded to two water molecules would predict. Instead, we find a collection of different arrangements where many DMSO molecules are connected to three water molecules by bond sharing while some are bonded to single water., along with others. We depict these clusters below.

The second important finding of this work is the strong dependence of the hydrogen bond lifetime on the composition. In fact, when we compute the coordination number of a methyl group from the methyl-methyl radial distribution function, we find that this coordination number bears a special significance as providing a quantitative estimate of the aggregation or crowding among the methyl groups. When the lifetime is plotted against this coordination number we find a strong correlation, thus providing a molecular level explanation of the strong composition dependence of hydrogen bond lifetime.

The organization of the rest of the paper is as follows. We first describe the details of the system simulated. Subsequently, we present an analysis of the composition dependence of the water-DMSO hydrogen bond strength through lifetime correlation functions. Subsequently, we show that hydrogen bond breaking is indeed coupled with fluctuation in the methyl group coordination. We develop an order parameter-based description of reaction free energy surface using the two well-known order parameters. Next, we use a two dimensional transition state theory to obtain the rate of hydrogen bond breaking and obtain a semi-quantitative agreement. Finally, we present a concise discussion of the results along with concluding remarks in connection with future problems.

## II.   System and simulation details

We consider water-DMSO binary mixtures by varying composition. This system has been studied extensively. Detailed simulations of water-DMSO mixture over a number of concentrations, from 0.001 % (one DMSO molecule in 1000 water molecules) to above 60% have been carried out. We note that water-DMSO mixture is known to exhibit multiple anomalies, both at low (15%) and intermediate (~35%) compositions. In case of water-DMSO mixture, we investigated only a limited range with the aim to elucidate the mechanism of strengthening proposed here. **Figure 2** shows the snapshot of the network structures are formed through H-bonding at different DMSO concentrations.

In this work, molecular dynamics simulations of water–DMSO binary mixture have been performed using GROningen MAchine for Chemical Simulations (GROMACS-5.0.7).[37] We employ the extended simple point charge (SPC/E) model of water. The DMSO molecules are modelled by the 4-site P2 model of Luzar and Chandler,[36] and recently employed by Rousseau et al.[21] The energy minimization has been performed using the steepest descent algorithm. We equilibrate the system in isothermal–isobaric ensemble (*NPT*; $T = 300$ K and $P = 1$ bar) and subsequently in canonical ensemble (*NVT*; $T = 300$ K) for 10 ns each. We carry out the final production runs in canonical ensemble (*NVT*; $T = 300$ K) for 5 ns using Berendsen thermostat.[38] The equations of motion were integrated using the leapfrog algorithm with a time step of 1 fs.[39] The data was dumped at a frequency of 100 fs. The simulations have been performed under periodic boundary conditions with a cut-off radius of 1.2 nm, and all the bonds were constrained using the LINCS algorithm.[40]

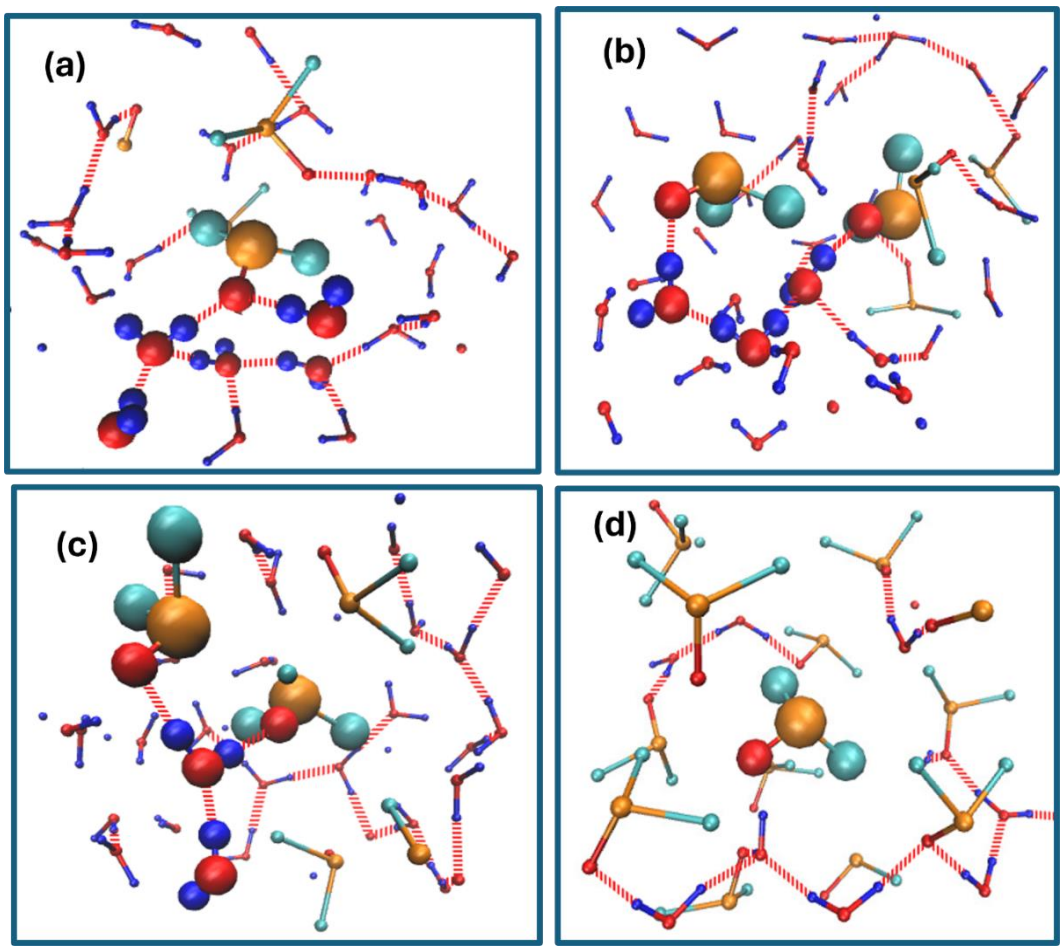

**Figure 2: Snapshot of the prevalent network structures that are formed through H-bonding. These have been arranged in a sequence of increasing composition, in going from (a) to (d), as follows. (a) DMSO molecules are completely solvated by the water molecule and form a network through H-bonding. This is the predominant structure at $X_{DMSO} < 0.20$. (b) &(c) Two DMSO molecules are connected through three water molecules or single water molecules respectively. These arrangements are observed at $X_{DMSO}=0.20-0.60$ (d) Above $X_{DMSO}= 0.60$, free DMSO molecules remain surrounded with self-aggregated DMSO. The red, blue, orange, and cyan spheres represent the oxygen atom, hydrogen atom, sulfur atom, and methyl group, respectively. The intermolecular hydrogen bonds are shown by red dashed lines.**

We would like to point out that our simulation produced many of the results that align with those discussed by Rousseau et al.[21] The main novel findings around the composition-dependent strength of water-DMSO hydrogen bonds and the analysis of the bond-breaking mechanism.

## III. Concentration induced strengthening of water-DMSO hydrogen bond

In order to obtain the hydrogen bond lifetime from simulation, we calculated hydrogen bond time correlation function $C_I(t)$, which is defined as allows,

$$C_I(t) = \frac{\langle h(0)h(t) \rangle}{\langle h(0)h(0) \rangle} \tag{1}$$

Here h(t) is a population parameter which attains a value '1' when a particular H-bond exists at time t, and '0' otherwise.

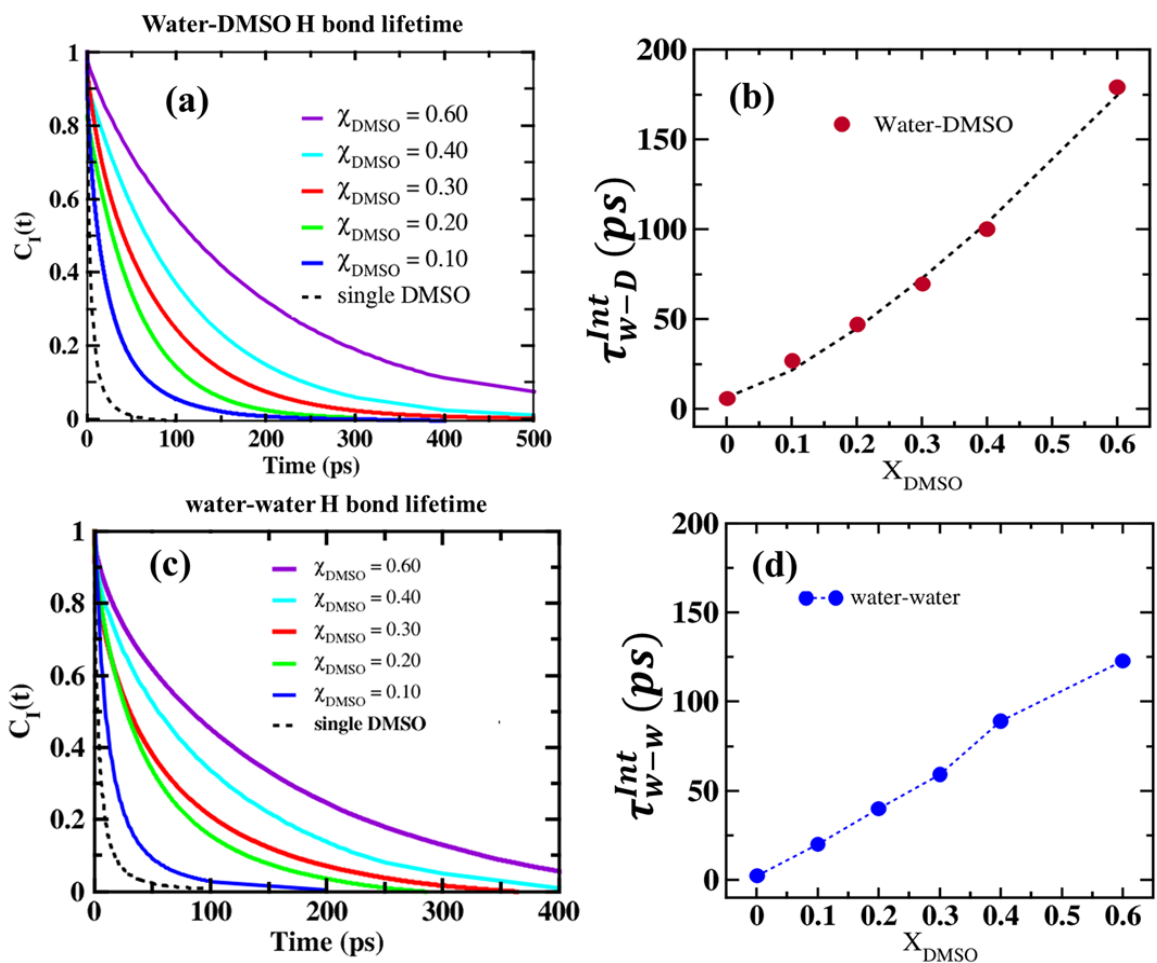

**Figure 3 (a) Water and DMSO hydrogen bond lifetime correlation functions $C_I(t)$ (b) hydrogen bond lifetime $\tau_{w-D}^{Int}$ as a function of composition. It shows nonlinear increases of lifetime with increase in DMSO concentration. Black dashed line shows power law fitting ($\tau_0 + a_1 \chi_{DMSO}^{a_2}$) curve, where $\tau_0$ = 6.04 ps, the lifetime at $X_{DMSO}$ =0.001, $a_1$= 335 ps, $a_2$=1.35, (c) water- water hydrogen bond lifetime correlation functions in single DMSO in water (black dashed line), 10 %**

**(blue line), 20 % (green line), 30% (red line), 40% DMSO-water (cayan line) and 60 % (violet line) binary mixture system. (d) Hydrogen bond lifetime $\tau_{w-w}^{Int}$ as a function of composition.**

Therefore, its time evolution corresponds to the structural relaxation of the hydrogen bonds, the relaxation times are provided in **table 1.**

**Table 1: Water-DMSO and water-water hydrogen bond lifetimes obtained from hydrogen bond time correlation function C$_I$(t). With increasing the composition of DMSO, $\tau_{w-D}^{Int}$ and $\tau_{w-w}^{Int}$ increases as shown in Figure 3(b), (d).**

| *Hydrogen bond lifetime of water-water (w-w) and water-DMSO (w-D) hydrogen bonds* | $\tau_{w-D}^{Int}$ (ps) | $\tau_{w-w}^{Int}$ (ps) |
|---|---|---|
| Single DMSO in water | 6.04 | 5.4 |
| 10 % DMSO | 26.8 | 20.2 |
| 20 % DMSO | 47.0 | 40.1 |
| 30% DMSO | 69.7 | 59.2 |
| 40% DMSO | 95.1 | 89.0 |
| 60% DMSO | 179.3 | 123.33 |

With increasing DMSO concentration, hydrogen bond relaxation times increases, which suggest attractive interaction between the associated hydrophobic alkyl groups results in the formation of a self-aggregate of DMSO molecules that strengthens the hydrogen bond.

A second definition of hydrogen bond, the continuous definition, denoted here by S$_c$(t) (sometimes referred to as S$_c$(t)), is often used to quantify the strength of the bond. In this definition, the bond is assumed to be broken once it crosses the bonding conditions. That is, the bond must be continuously present. Mathematically, it is given by

$$S_c(t) = \frac{\langle h_c(0) h_c(t) \rangle}{\langle h_c(0) h_c(0) \rangle} \quad (2)$$

Here $h_c(t) = 1$ when the tagged pair of molecules remain hydrogen bonded till time t, and '0' otherwise. The time constants obtained from $S_C(t)$ have been provided in the Supplementary Material. As is well-known, the decay of the lifetime function is much faster than that of the intermittent function. Sometimes the former behaves differently than the latter in its dependence on environmental variables, like composition here.

## IV. Fluctuation induced hydrogen bond breaking

Luzar and Chandler investigated the structural properties,[36] H-bond distribution, and H-bond dynamics of water in the aqueous solution of DMSO ($X_{DMSO}$ = 0.35) by neutron diffraction experiments.[41] They discovered that in solutions containing aqueous DMSO, the most significant anomalies are observed around $X_{DMSO}$ ~ 0.35. It is to be noted that hydrogen bond stoichiometry dominates at $X_{DMSO}$ = 0.35, favouring DMSO-water complexes in 1DMSO:2H$_2$O ratio. [21,36,41–43]

In order to understand and quantify the role of hydrophobic interactions, we introduce a new quantity $n_{me}$ which is defined as the number of methyl group with in the cut off distance 4.5Å of the methyl group of tagged DMSO molecules. We plot the time evolution of $n_{me}$ in **Figure 4**. Large fluctuations in $n_{me}$ are observed prior to the disruption of the hydrogen bond.

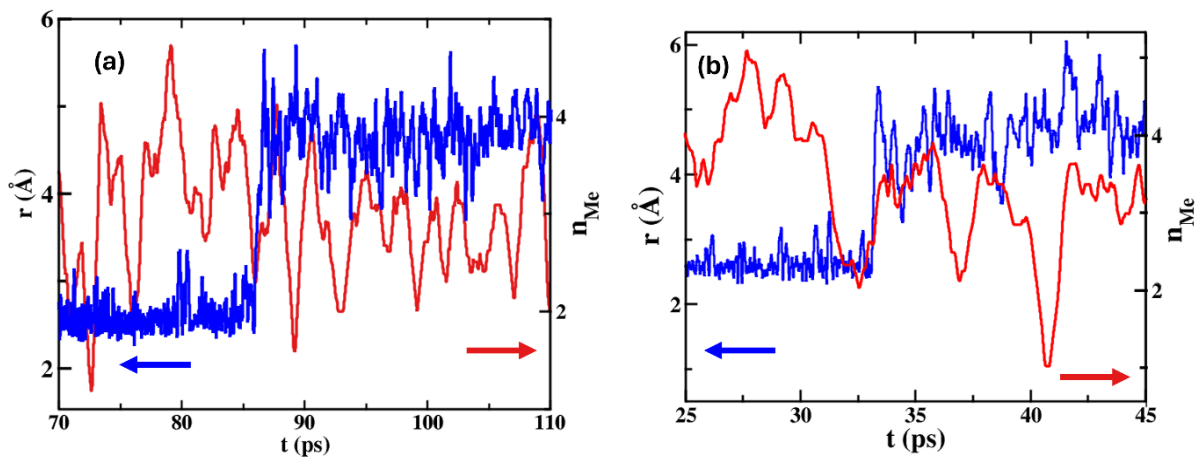

**Figure 4: Donor-Acceptor separation distance (r) and $n_{me}$ are plotted against the time for two tagged hydrogen-bonded pairs. Significant fluctuations in $n_{Me}$ occurs before the hydrogen bond disruption.**

We further quantify the correlation between $n_{me}$ and r by calculating the Pearson correlation coefficient (ρ) between them [Eq. (3)].

$$\rho = \frac{\langle \delta n_{me} \delta r \rangle}{\sqrt{\langle \delta n_{me} \rangle^2 \langle \delta r \rangle^2}} \quad (3)$$

In Eq. (3). $\delta n_{me}, \delta r$ denotes deviation of $n_{me}$, r with respect to an average calculated over 150 ps trajectory (time step 10 fs). A positive *ρ* implies correlated trajectories, while a negative value indicates anti-correlation. The two quantities are uncorrelated if *ρ* is 'zero'. Here we observe that r and $n_{Me}$ are anti correlated. (Pearson's correlation coefficients are found to be -0.45, -0.32).

## V. Effects of crowding

In order to understand hydrophobic interaction-induced H bond stability between water-DMSO, we further calculate the radial distribution function (rdf) between the methyl groups ($CH_3$- or Me-) of distinct DMSO molecules that gives us an idea of the presence of other DMSO molecules in the neighbouring shell. In **figure 5**, $4\pi r^2 \rho g(r)$ is plotted as a function of r for five different compositions. Here ρ is the number density of the DMSO, and *r* is the separation between the two species.

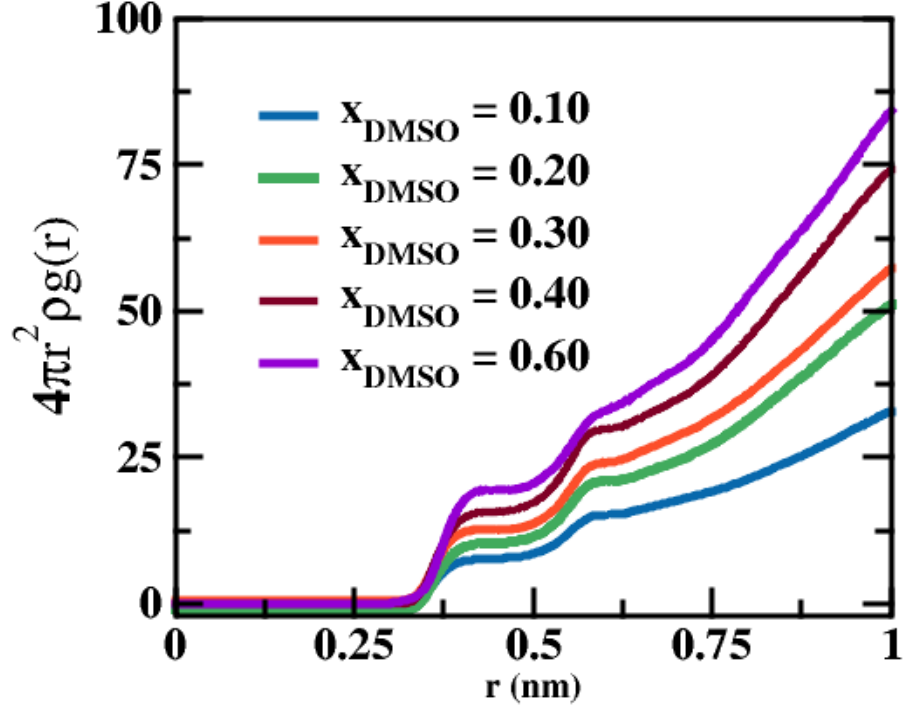

**Figure 5:** $4\pi r^2 \rho g(r)$ is plotted against r, the separation distance between the two methyl- group. The first neighbouring shell extends up to ~ 5 Å. Here g(r) the radial distribution function of methyl groups of two distinct DMSO molecule, ρ is the number density of the DMSO, and r is the separation between the two species.

It also provides us with a cut-off value of **5 Å** as a radius of the first neighbouring shell and within that cutoff, we calculate the average number of methyl groups surrounding a central methyl group by using the following equation.

$$CN_{me} = 4\pi\rho \int_0^{r_{min}} dr g(r) r^2 \qquad (4)$$

where $g(r)$ is the radial distribution function between the two-methyl group, $r_{min}$ is the position of the first minimum of $g(r)$.

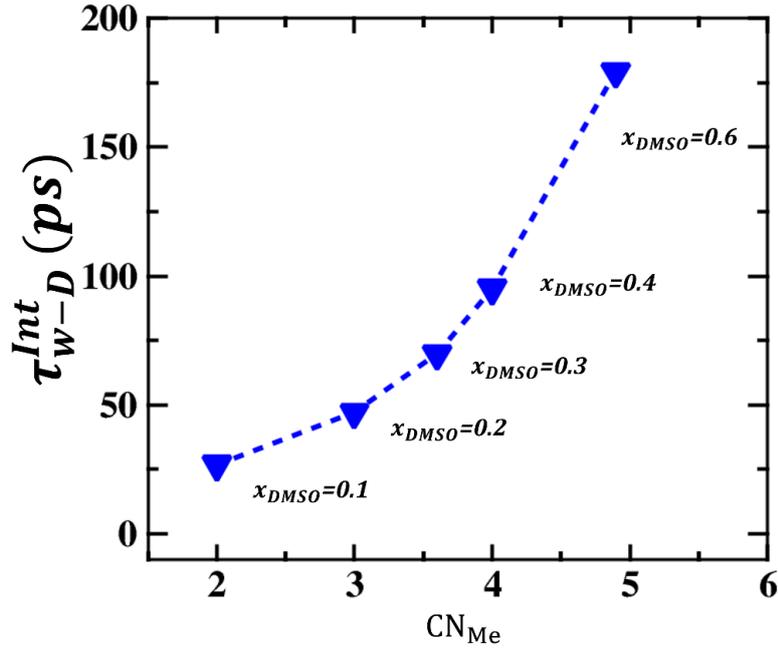

**Figure 6 : The water-DMSO hydrogen bond lifetime [ $\tau_{w-D}^{Int}$ (ps)] is plotted against $CN_{Me}$, for five different concentrations of DMSO such as 10%, 20%, 30%, 40% and 60%. [ $\tau_{w-D}^{Int}$ (ps)] shows nonlinear increment with $CN_{Me}$. This non-linear dependence of lifetime on the $CN_{Me}$ confirms the role of hydrophobic attraction in the slowdown of lifetime**

We plot $\tau_{w-D}^{Int}$ as a function of $CN_{Me}$, for five different concentrations of $X_{DMSO}$ such as 0.10, 0.20, 0.30, 0.40 and 0.60. Note that as the crowding increases sharp increases of $\tau_{w-D}^{Int}$ is observed. It is evident from **Figure 6** that $CN_{Me}$ shows a positive correlation with DMSO concentration. Plotting the lifetime of water-DMSO hydrogen bonds against this coordination number unveils a distinct non-linear relationship, suggesting the contribution of hydrophobic attraction to the slowdown of hydrogen lifetime.

## VI. Free energy surface

We consider the following geometrical definition to determine the hydrogen bond breaking, namely (i) Two molecules are hydrogen-bonded when their inter-oxygen distance (r) is shorter than the threshold value of 3.5Å and (ii) The angle ($\theta$) O–H···O must be lesser than the threshold value of 30°. Time evolution of r and $\theta$ are shown in figure.

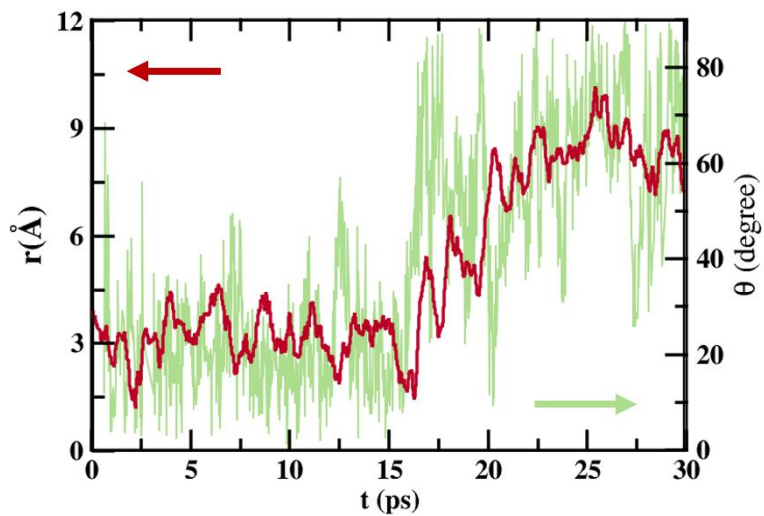

**Figure 7: Time evolution of the donor acceptor distance r (brown line), and $\theta$ (green line)**

Further, we construct a potential mean forced based free energy surface for water–water hydrogen bond (a), and water-DMSO hydrogen bond (b) with respect to these two order parameters, r and θ for 0.001% and 30% water-DMSO binary mixture. **Figure 8** shows the mean force-based free energy surface of water-DMSO hydrogen bond breaking for $X_{DMSO}$ = 0.001 and **Figure 9** shows the free energy surface for water-water hydrogen bond breaking [(a), (b)] and water-DMSO hydrogen bond breaking [(c),(d)] for $X_{DMSO}$ = 0.30. The minima around r ~ 3Å and θ< 30⁰ signifies the hydrogen bonded state.

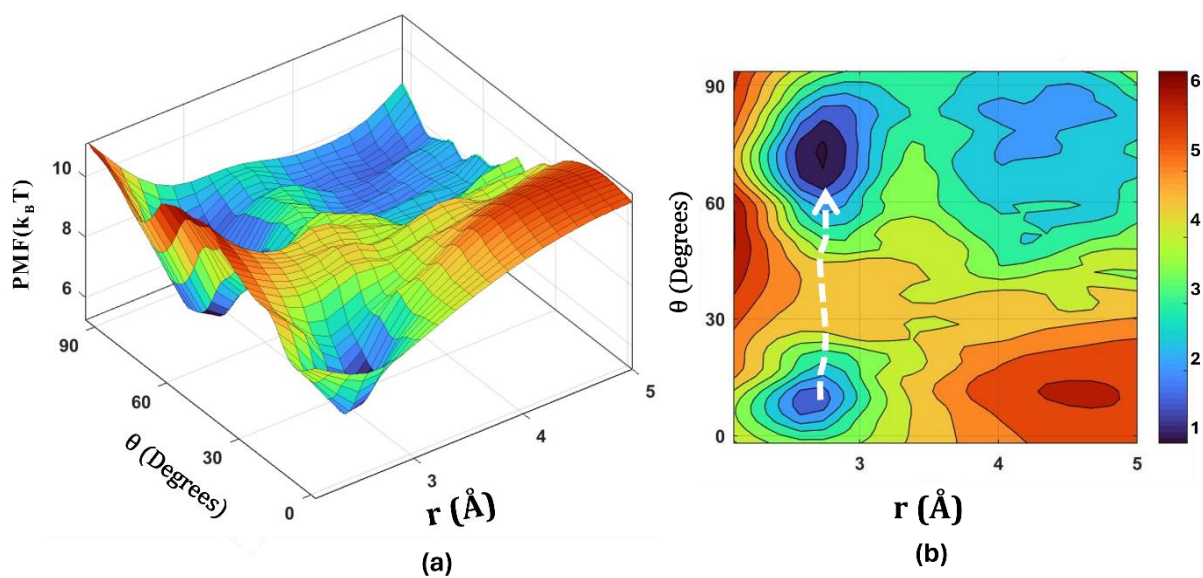

(a)  (b)

**Figure 8: (a) The potential of the mean force-based free energy surface of water-DMSO hydrogen bond breaking and (b) two-dimensional projection of the free energy surface. The blue patches represent free energy minima.**

**Figures 8(b)** and **9(b)** suggest two distinctly different bond breaking mechanism at two different local environments. For single DMSO molecules first free energy minima appears (r < 3Å and θ<30º) and second minima r ~2.5- 3Å and θ~70º-80º ). The h bond breaking is initiated through the rotational relaxation as depicted in figure 8 (a). On the other hand for $X_{DMSO} = 0.30$, hydrophobic methyl groups results in self-aggregation of DMSO molecules (quantified by the average coordination number of individual methyl groups), which "cages" both rotational and linear motions of molecules involved and, in effect, strengthens the water-DMSO hydrogen bond network.

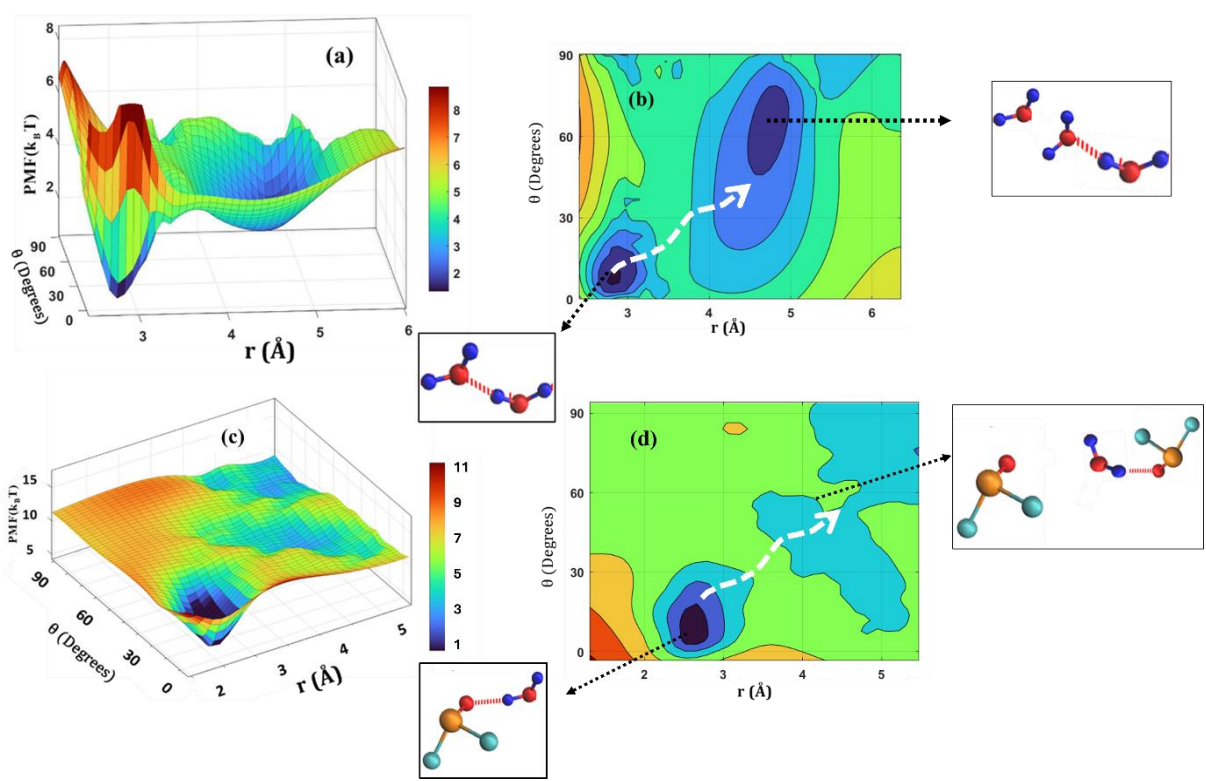

**Figure 9. The potential of the mean force-based free energy surface of (a) water-water hydrogen bond breaking and (c) water-DMSO hydrogen bond breaking. The energy barriers for hydrogen bonding in water-water and water-DMSO are observed to be 3.5 $k_BT$ and 4.5 $k_BT$, respectively.**

**Two-dimensional projection of the free energy surface with respect to two order parameters (r, θ), (b) water-water hydrogen bond breaking and (d) water-DMSO hydrogen bond breaking. The blue patches represent free energy minima. H-bonded and unbonded state of the tagged molecule are provided in the inset.**

## VII. Hydrogen bond breaking rate calculation

Utilizing the free energy surface, we calculate the water-water and water-DMSO hydrogen bond breaking rate by using 1D and 2D transition state theory. We provide a calculation of the bond breaking rate using one-dimensional and two-dimensional transition state theory. The calculation of the reaction rate requires two parameters, the free-energy barrier $\Delta G$, and the reactant well frequency $\omega_R$.

$$k_{TST} = \frac{\omega_R}{2\pi} \exp\left(-\frac{\Delta G}{k_B T}\right) \tag{5}$$

In Eq.(5), $k_B$ represents the Boltzmann constant, and T signifies the absolute temperature, $\Delta E$ represents the energy barrier, which is 3.5 $k_B T$ and 4.5 $K_B T$ for water-water hydrogen bond break and . The reactant well frequency is determined by fitting the potential energy surface with a parabolic function.

Two-dimensional transition state theory rate constant can be expressed in terms of one-dimensional transition state theory expression [Eq.(5)] as follows

$$k_{TST}^{2D} = \frac{\omega_Y^w}{\omega_Y^B} k_{TST} \tag{6}$$

Here $\omega_Y^w$ and $\omega_Y^B$ are the reactant well frequency and Barrier frequency along the Y direction. In order to get the well frequency and barrier frequency, we fit the well and the barrier with parabolic function and inverse parabolic function respectively. [Barrier frequency and well frequency are provided in supporting information]. The bond breaking lifetimes are given below in the tabular form.

**Table 2: Water-water and water-DMSO hydrogen bond breaking lifetimes obtained from 1D TST , 2D TST theory [Eq. (5) and (6)] and simulation.**

| System | H bond type | 1D TST ps | 2D TST ps | Simulation ps |
|---|---|---|---|---|
| $X_{DMSO}= 0.001$ | Water-Water | 3.1 ± 0.32 | 3.6 ± 0.21 | 5.4 |
| | Water-DMSO | 3.9 ± 0.30 | 5.6 ± 0.40 | 6.04 |
| $X_{DMSO}= 0.30$ | Water-Water | 22.32 ± 2.0 | 35.48 ± 2.5 | 59.2 |
| | Water-DMSO | 33.12 ± 3.6 | 55.60 ± 3.1 | 69.7 |

It is observed that time constant of the hydrogen bond breaking increases as we go from 1D TST to 2D TST. The transition state theory establishes a maximum value for the rate constant by neglecting frictional influences. The rate constant obtained from the TST is in good agreement with the simulation result.

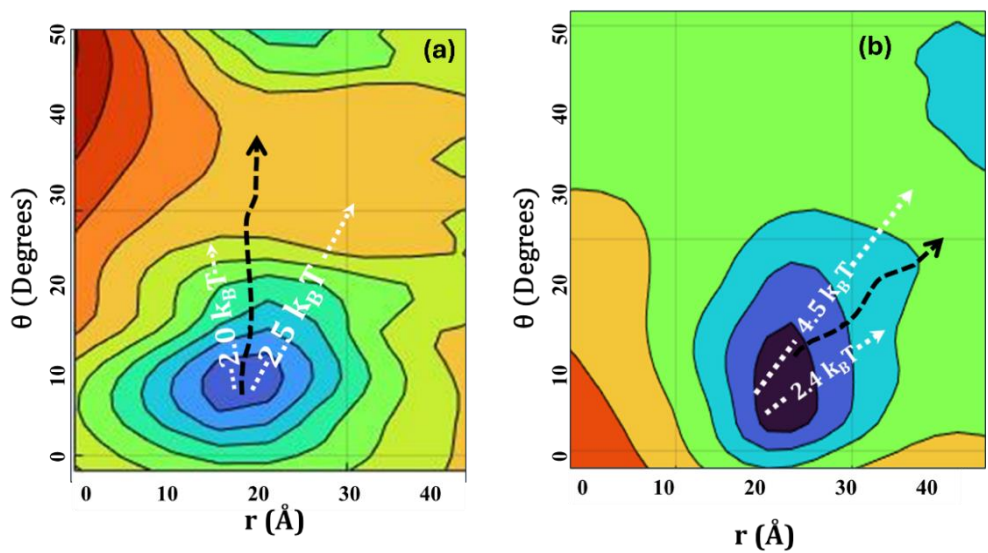

**Figure 10: Contour plot water-DMSO h-bond breaking. We present only reactant well region. The concerned collective variables are the donor acceptor separation distance (r) and the O–H···O bond angel $\theta$. (a) In case of $X_{DMSO}$ = 0.001, h-bond breaking is initiated through rotational relaxation (b) In case of $X_{DMSO}$ = 0.03 h-bond breaking is initiated through translational relaxation. It is observed that the energy barrier for the rapid escape of water molecule from bonded state is 2.0 $k_BT$ and 2.4 $k_BT$ for $X_{DMSO}$ = 0.001 and $X_{DMSO}$ = 0.03 respectively. It suggests, water molecule can reach the top of the basin and then return back to its minima. These unsuccessful attempts of escape which is prominent from a downward gradient in the free energy surface. Energy barriers correspond to the ultimate escape is 2.5 $k_BT$ and 4.5 $k_BT$ for $X_{DMSO}$ =0.001 and $X_{DMSO}$ = 0.03 respectively.**

The calculation of the hydrogen bond lifetime essentially involved the calculation of mean first passage time. It depends on reaching to a particular space in configuration space, as shown in Figure 10 (a), (b). However, the free energy surface has a downward slope, which suggests it can return back to the free energy minima, as shown by the unsuccessful attempt to escape from the minima. According to the definition, the Continuous function ($S_c(t)$) doesn't allow the escape; on the other hand, the Intermittent function ($C_I(t)$) allows return and, ultimately, escape. According to the chemical reaction kinetics theory we can model this rate of escape as an absorbing barrier beyond the free energy minima as described in Kramers' theory. According to transition state theory, one can place an absorbing barrier at the top of the basin. Therefore, ratio of the time constant of the two-process describe by $C_I(t)$ and $S_c(t)$ given by the ratio of the exponential factor of activation energy of the two processes. For the case of $X_{DMSO}$ = 0.001 and $X_{DMSO}$ = 0.30, free energy barrier separating these two points are 0.5 $K_BT$ and 2.1 $K_BT$, which leads corresponding ration of their exponential factor as 1.6 and 8.2 respectively. Whereas the ration of $\tau_{W-D}^{int}$ and $\tau_{W-D}^{C}$ for $X_{DMSO}$ = 0.001 and $X_{DMSO}$ = 0.30 obtained from simulation are 4.12 and 13.2 respectively [**Table 3**]. The agreement of this value is not accurate, because of the transition state approximation but it well explains much larger Intermittent lifetime that than of the continuous lifetime of hydrogen bond breaking.

**Table 3. In the first column, the ratio of C$_I$(t) and Sc(t) for X$_{DMSO}$ = 0.001 and X$_{DMSO}$ = 0.30 obtained from simulation results are provided. In the second column predicted ration of two processes obtained from the FES are presented.**

| X$_{DMSO}$ | $\frac{\tau_{W-D}^{Int}(t)}{\tau_{W-D}^{C}(t)}$ | Predicted ratio from FES |
|---|---|---|
| **0.001** | 4.12 | 1.6 |
| **0.30** | 13.8 | 8.2 |

**VIII. Conclusions**

The enhanced stability of self-aggregates has always been attributed to the existence of multiple interactions which act together to make these systems robust and long-lived. Because of the complexity of these systems, a theoretical study has always proven difficult. The presence of water-water and water-DMSO hydrogen bonds create a complex network which admits of various quasi-stable local structures. The occurrence and prevalence of these local structures where hydrophobic interactions between the methyl groups of DMSO needs also to be accommodated create study of these structures a fascinating exercise, particularly in computer simulations.

The present study is different from a large number of earlier studies on the same water-DMSO binary mixture in its focus on understanding the composition dependence of the hydrogen bond lifetime. We attempt to understand the relatively rapid emergence of slow dynamics as the DMSO composition in the mixture is increased. The water-DMSO increases by more than order of magnitude from very low concentration to that at 30%. Somewhat to our surprise, even the water-water h-bond lifetime also increases by almost the same amount.

The above results are somewhat intriguing. We attributed the emergence of this slowness to a hydrophobic interaction between the methyl groups of DMSO. The resulting formation of the structure is found to involve a bridge between two DMSO by a single water molecule or by a chain of three water molecules, with the latter being more predominant at low DMSO concentrations, with the former component more dominant at around 50% DMSO composition. We note that when two DMSO molecules are bridged by three water molecules. In each case, the DMSO oxygen accepts two hydrogen bonds, as was proposed by Luzar and Chandler. The simultaneous existence of such structures pushes the maximum in the composition dependence of viscosity to near 40% (around 38%, to be precise), higher than 33% expected from estimates usually surmised from the 1DMSO:2H$_2$O ratio.

In confirmation of the pivotal role of the hydrophobic attraction between the methyl groups of DMSO, we calculate the coordination number CN$_{Me}$ of each methyl group as a function of composition. Clearly CN$_{Me}$ is an increasing function of DMSO concentration. When the lifetime of water-DMSO hydrogen bond is plotted against this coordination number, a clear non-linear dependence of lifetime on the coordination number (Figure 6) is revealed which confirms the role of hydrophobic attraction in the slowdown of lifetime and also of the liquid mixture.

The crucial role of hydrophobic interaction in strengthening the water-DMSO hydrogen bond is further revealed by analyses of the trajectories, as plotted in Figure 4 where we demonstrate the correlation between the increase of Me-Me distance that occurs via natural thermal fluctuation leads to the breaking of the said hydrogen bond. This correlation between fluctuation and bond breaking events has been observed in many trajectories. [provided in supplementary material] This agrees with our hypothesis that hydrophobic attraction between the methyl groups.

There are still several unanswered questions. For example, while the lifetime of the hydrogen bond continues to increase with DMSO concentration, viscosity exhibits a sharp maximum at a concentration $X_{DMSO}$=0.38. The subsequent decrease in the value of viscosity is rather sharp. This sharp decoupling of hydrogen bond lifetime could arise from certain cross-correlations between water and DMSO contributions to the stress-stress time correlation function that requires further work. We note that the associated second characterization of hydrogen bond lifetime through the intermittent survival correlation function, $S_I(t)$ does show a non-monotonic dependence on composition. These facts seem to suggest that viscosity is controlled by the short time dynamics.

The intriguing dynamical properties of water-DMSO binary mixture might not be specific to this mixture alone, but some aspects could be present in such systems as water-ethanol and water-ethanol mixtures. This interplay between hydrogen bond and hydrophobic interaction could be a general phenomenon in many aqueous mixtures. This problem certainly deserves further study.

## Supplementary Material

See the supplementary material for the following: (i) forcefield parameters, (ii) continuous hydrogen bond lifetime. (iii) trajectories of tagged molecules (iv) well and barrier frequencies.

## Acknowledgement

We thank Mr. Shubham Kumar and Mr. Subhajit Acharya for useful discussions. B.B. thanks the Science and Engineering Research Board (SERB), India, for the National Science Chair Professorship and the Department of Science and Technology, India, for partial research funding. S.M. thanks IISc. for a research fellowship. We thank Supercomputer Education and Research Centre, IISc for computational facilities.

# Supplementary Material

# Composition Dependence of Hydrogen Bond Lifetime in Water-DMSO Binary Mixtures: The Role of Hydrophobic Interaction
## Sangita Mondal and Biman Bagchi*

*SSCU, Indian Institute of Science, Bangalore 560012, India.*

*\*Email: bbagchi@iisc.ac.in; profbiman@gmail.com*

Contents

S1. Force field parameters.

S2. Continuous hydrogen bond time correlation function

S3. Trajectories of tagged molecules

S4. Barrier and well frequencies

In this Supplementary Material part, we have presented further analyses with numerical results to support and supplement the main results of the text.

# S1. Force field parameters of water, DMSO and Ions

**Table S1. Parameter set for water adapted from SPC/E Force field[1]**

|   | σ(Å) | ϵ(Kj/mol) | Charge (e) |
|---|---|---|---|
| O | 3.169 | 0.6502 | -0.8476 |
| H | ……. | …….. | +0.4238 |

**Table S2. Parameter set for DMSO adapted from e 4-site P2 model of Luzar and Chandler[2] Force field**

|   | σ(Å) | ϵ(Kj/mol) | Charge (e) |
|---|---|---|---|
| CH$_3$ | 3.80 | 1.23 | +0.16 |
| S | 3.40 | 1.0 | +0.139 |
| O | 2.80 | 0.30 | -0.459 |

# S2. Continuous hydrogen bond lifetime

The continuous definition, denoted here by $S_C(t)$ (sometimes referred to as $S(t)$), is s frequently employed to measure the hydrogen bond's. In this definition, the bond is assumed to be broken once it crosses the bonding conditions. That is, the bond must be continuously present. Mathematically, it is given by Eq. (2). Where $h_c(t) = 1$ when the tagged pair of molecules remain hydrogen bonded till time t, and '0' otherwise.

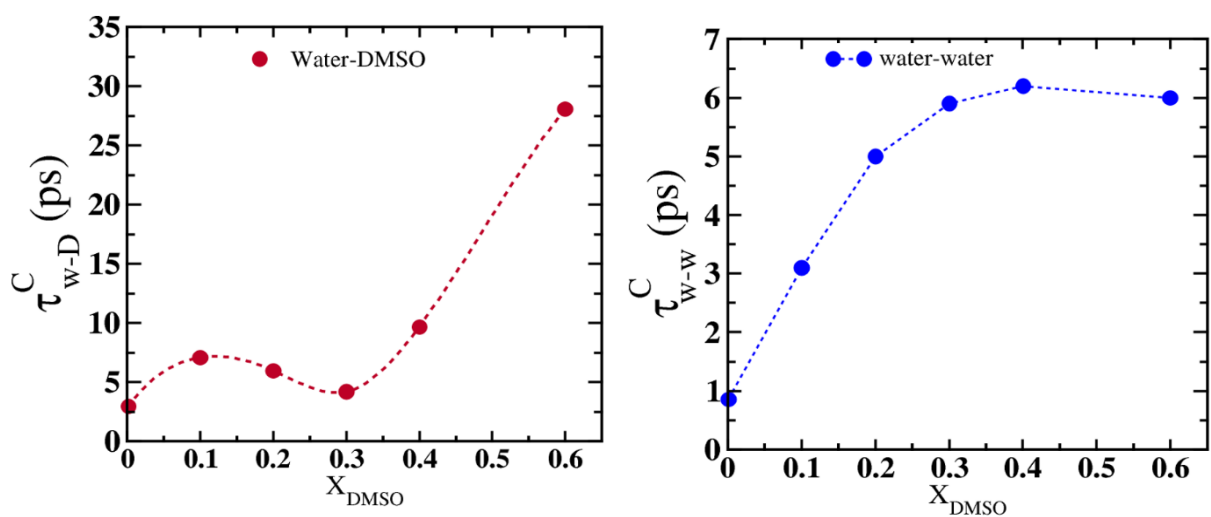

**Figure S1. Continuous hydrogen bond lifetime (a)** $\tau^C_{w-D}$ **,(b)** $\tau^C_{w-w}$ **as a function of composition.**

It is observed that the water-DMSO continuous hydrogen bond lifetime $\left(\tau^C_{w-D}\right)$ is non monotonically increasing with increase in DMSO concentration, where as water-water continuous hydrogen bond lifetime $\left(\tau^C_{w-w}\right)$ shows a monotonic increase, with an increase in DMSO composition.

## S3. Trajectories of tagged molecules

Here we have shown the trajectories of the molecule. We have plotted the time evolution of $n_{Me}$ (the number of the methyl groups within the cut-off distance of 4.5Å of the methyl group of tagged DMSO molecules) and the bond donor-acceptor distance r. We observe large fluctuations in $n_{me}$ prior to the disruption of the hydrogen bond.

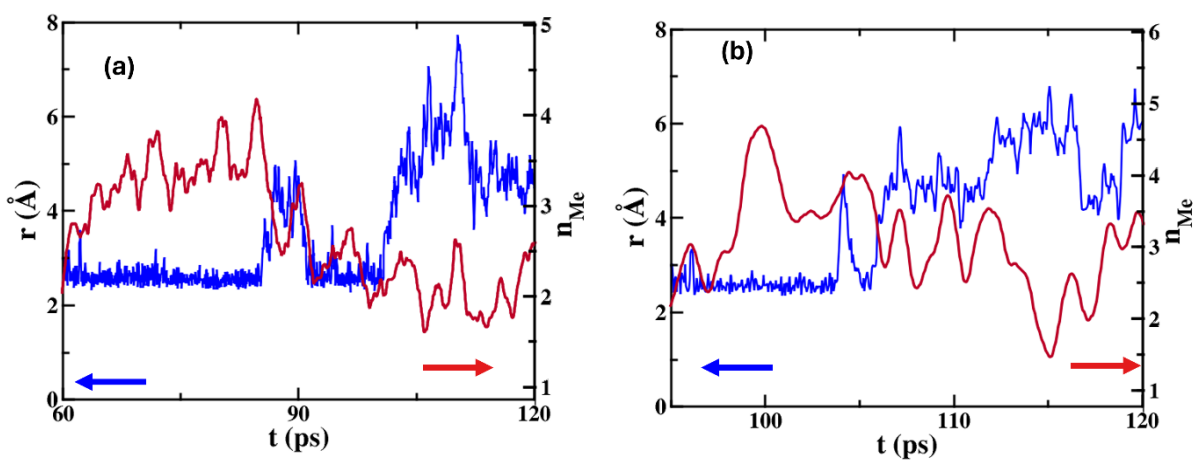

**Figure S2. Donor-Acceptor separation distance (r) and $n_{me}$ are plotted against the time for two tagged hydrogen-bonded pairs.**

## S4. Barrier and well frequencies

The energy barrier, barrier frequency, and well frequencies used in the hydrogen bond breaking rate are provided in tabular form.

**Table S3**. **Energy barrier, barrier frequency, and well frequencies**

| Composition | h-bond type | Energy barrier ($k_BT$) | $\omega_R$ (ps$^{-1}$) | $\omega_Y^w$ (ps$^{-1}$) | $\omega_Y^B$ (ps$^{-1}$) |
|---|---|---|---|---|---|
| $X_{DMSO} = 0.001$ | Water-water | 2.4 | 21.4 | 0.008 | 0.01 |
|  | Water-DMSO | 2.5 | 19.3 | 0.0128 | 0.018 |
| $X_{DMSO} = 0.30$ | Water-water | 3.5 | 17.2 | 0.014 | 0.017 |
|  | Water-DMSO | 4.5 | 17.2 | 0.0048 | 0.008 |